# Structural Determination vs Stochastic Events: Relaxation Dynamics in a Supercooled Liquid

Shivam Mahajan[a], Peter Harrowell* [b], Massimo Pica Ciamarra^ [a,c]

[a] Division of Physics and Applied Physics, School of Physical and Mathematical Sciences, Nanyang Technological University, Singapore 637371, Singapore

[b] School of Chemistry, University of Sydney, Sydney New South Wales, 2006 Australia

[c] CNR-SPIN, Dipartimento di Scienze Fisiche, Università di Napoli Federico II, I-80126 Napoli, Italy

Abstract

Structural relaxation in supercooled liquids is influenced by the initial configuration, yet its microscopic evolution exhibits substantial variability. Here, we investigate the interplay between structural constraints and dynamical stochasticity by examining fluctuations of relaxation dynamics across and within isoconfigurational ensembles. We show that fluctuations among trajectories originating from the same configuration constitute a substantial fraction of the total dynamical fluctuations over a broad range of temperatures. At the collective level, the distribution of relaxation times is well described by a stochastic model involving a sequence of effective relaxation events, whose number decreases upon cooling. At the microscopic level, dynamical fluctuations exhibit spatial correlations that grow upon cooling, revealing a spatial organization of relaxation even when the initial configuration is fixed. Remarkably, the characteristic length associated with these fluctuations closely follows the length scale inferred from the stochastic first-passage model, establishing a connection between the spatial organization of dynamical fluctuations and the temporal statistics of collective relaxation. Together, these results suggest that structural relaxation is governed by an interplay between configuration-dependent constraints and stochastic dynamical events: the initial structure influences when and where relaxation occurs, without uniquely determining its microscopic realization.

*peter.harrowell@sydney.edu.au

^massimo@ntu.edu.sg

# 1 Introduction

The microscopic characterization of cooperative dynamics in a supercooled liquid inevitably runs up against the difficulty of making intelligible the complex conflation of collective strains, coordinated reorganizations and thermal noise that, together, produce structural relaxation. One strategy to deal with this problem has been to replace the complex reality with a simple but plausible representation of the dynamics. Examples of this approach include single particle hops [1,2], diffusing defects [3–5], or continuum elastic-plastic models [6] . The result in each case is a tractable dynamics but a dynamics whose relation to reality rests on speculation. An alternative approach, one that avoids the insertion of an idealized dynamic model, is to forego any explicit account of dynamics and, instead, to focus on resolving structures that act to restrain particle motion and structures that permit movement of some sort. The resulting study of dynamic heterogeneities in supercooled liquids and their connection to the disordered structures that sustain them represents a major theme in the field over the last 30 years. The propensity for motion has proved to be a useful tool in the quest for a causal link from structure to dynamics in a supercooled liquid [7–9]. The idea is to select a local measure of relaxation (e.g. particle displacement, overlap with initial position, fraction of initial neighbors retained, etc.) and then to average this quantity over an ensemble of trajectories all starting from the same initial configuration. Across these trajectories, the initial structure is held fixed while the dynamical realization is varied, for example through different initial momenta in deterministic dynamics or different stochastic realizations in thermostatted settings. This ensemble is called the isoconfigurational ensemble, and the associated average relaxation function is the propensity at a given time [7] . Once the dependence on the particular dynamical realization has been averaged out, any spatial variation in the propensity, so the argument goes, must arise from structural determination of the dynamics by features of the initial configuration.

The propensity has come to represent something of a gold standard in isolating that aspect of the relaxation dynamics that is influenced by structure [8,10–14]. Implicit in this success is the fact that there do exist contributions to the fluctuations in dynamics that are not directly linked to the structure and so are averaged out over the isoconfigurational ensemble. These fluctuations arise from differences between trajectories originating from the same initial configuration, and we refer to them as *isoconfigurational fluctuations*. Our goal in this paper is to characterize these fluctuations, quantify their importance relative to fluctuations between different initial configurations, and investigate their temporal statistics and spatial organization. In particular, we examine whether isoconfigurational fluctuations retain any connection with the initial structure, despite arising between trajectories that share the same initial configuration.

The paper is organized as follows. In Section 2, we introduce the model system and define the overlap function used to characterize structural relaxation. In Section 3, we quantify the contributions of isoconfigurational and configurational fluctuations to dynamical variability, at both the collective and individual-particle levels. In Section 4, we investigate the statistics of first-passage times and introduce a stochastic model to describe the observed fluctuations. In Section 5, we examine the spatial correlations of dynamical fluctuations and compare the resulting correlation lengths with the characteristic length scale inferred from the stochastic model. Finally, we summarize our main findings in Section 6.

# 2 Model and Algorithm

In this paper we have used a 2d binary Lennard-Jones mixture introduced by Brüning et al. [15]. The potential is the same as used by Kob and Andersen [16] to model the glass transition in 3d but the composition is adjusted from $A_{80}B_{20}$ to $A_{65}B_{35}$ to avoid crystallization, as discussed in [15]. A linear smoothing ensures the potential and the force vanish at the cutoff distance. The density is fixed at $\rho = 1.2$, and the dynamics is investigated via Nosé-Hoover molecular dynamics simulations in the NVT ensemble. We consider systems with $N = 1000$, 2000 and 4000 particles. At each temperature, we prepare 200 independent equilibrated configurations for N = 1000, and 100 for each of the larger system sizes, verifying equilibration through the absence of aging in the relaxation dynamics. For each equilibrated configuration, we then generate at least 50 isoconfigurational runs. To verify the generality of our results, we have also performed simulations of a three-dimensional Kob and Andersen mixture, which we report in Appendix A.

Throughout the paper, we use cage-relative displacements to suppress the contribution of long-wavelength collective motion [17–21] . For each particle $i$, we identify its $N_i$ nearest neighbors in the initial configuration using a cutoff distance of $1.64\rho^{-1/2}$, just below the first minimum of the radial distribution function. The cage-relative displacement is

$$\Delta\boldsymbol{r}_i^{\mathrm{cr}}(t) = \boldsymbol{r}_i(t) - \boldsymbol{r}_{\mathrm{i}}(0) - \frac{1}{N_i}\sum_j \left[\boldsymbol{r}_j(t) - \boldsymbol{r}_j(0)\right] \tag{1}$$

where the last sum extends over all neighbors. We then define the overlap of particle $i$ at time $t$ with the time evolved position of the cage made up of its initial neighbors as

$$o_i(t) = \Theta\left(a - |\Delta\boldsymbol{r}_i^{\mathrm{cr}}(t)|\right), \tag{2}$$

which equals one if the cage-relative displacement is smaller than the cutoff $a = 0.3\ \rho^{-1/d}$, and zero otherwise. The investigated temperature range, $0.36 \leq T \leq 1$, extends down from the onset temperature [22], $T_{on} \simeq 1$, and encompasses our estimation of the mode-coupling critical temperature, $T_c \simeq 0.39$.

# 3 Isoconfigurational fluctuations are large

To assess how much dynamical variability remains after fixing the initial configuration, we examine fluctuations in the overlap dynamics at two levels. At the collective level, we consider fluctuations of the system-averaged overlap, which reflect both the magnitude and spatial correlations of particle-level fluctuations. At the microscopic level, we examine the variability of individual particle overlaps, independently of their spatial correlations. These complementary perspectives allow us to distinguish the importance of isoconfigurational fluctuations in collective relaxation from their contribution to the unpredictability of individual particle dynamics.

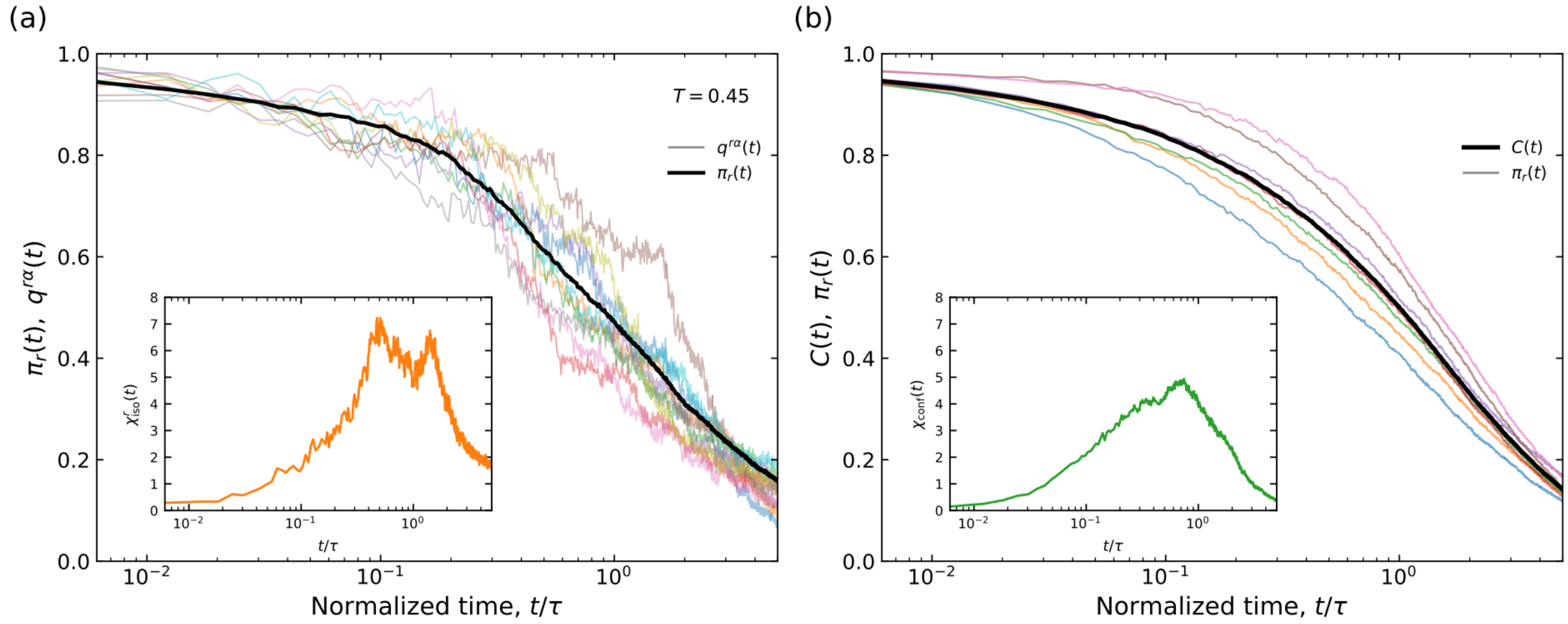

*Figure 1. (a) A selection of individual relaxation function $q^{r\alpha}(t)$ from a single initial configuration (pale lines), representing ten random relaxing trajectories. The black line is the collective propensity $\pi_r(t) = \langle q^{r\alpha}(t)\rangle_\alpha$ for the same initial configuration. The fluctuations of the individual relaxation functions define the isoconfigurational susceptibility for the considered configuration, $\chi^r_{iso}(t) = N\,Var_\alpha[q^{r\alpha}(t)]$, we illustrated in the inset. (b) A selection of ten collective propensities $\pi_r$ (pale lines), associated with different initial configurations. The black line is the standard relaxation function $C(t) = \langle \pi_r(t)\rangle_r$. The fluctuation of the collective propensity defines the configurational susceptibility, $\chi_{conf}(t) = N\,Var_r[\pi_r(t)]$, we illustrate in the inset. All data refer to T = 0.45, N = 1000.*

## 3.1 Collective fluctuations

For a trajectory $\alpha$ starting from a given initial configuration $r$, we define the collective overlap as the spatial average of the individual particle overlaps,

$$q^{r\alpha}(t) = \frac{1}{N}\sum_{i=1}^{N} o_i^{r\alpha}(t) \tag{3}$$

Averaging over the isoconfigurational ensemble of trajectories originating from the same configuration gives the collective propensity,

$$\pi_r(t) = \langle q^{r\alpha}(t)\rangle_\alpha = \frac{1}{N}\sum_{i=1}^{N}\langle o_i^{r\alpha}(t)\rangle_\alpha = \frac{1}{N}\sum_{i=1}^{N}\pi_{ri} \tag{4}$$

where $\pi_{ri} = \langle o_i^{r\alpha}(t)\rangle_\alpha$ is the standard particle propensity [16,23] of particle $i$.

Further averaging over initial configurations yields the standard relaxation function,

$$C(t) = \langle \pi_r(t)\rangle_r = \langle\langle q^{r\alpha}(t)\rangle_\alpha\rangle_r \tag{5}$$

We define the relaxation time $\tau$ through $C(\tau) = 1/2$.

In 2003, Lačević et al. [24] introduced the susceptibility $\chi_4(t)$ defined as the variance of the relaxation function over both the initial configuration and trajectories,

$$\chi_4(t) = N\,\mathrm{Var}_{r\alpha}[q^{r\alpha}(t)] = N\left\langle\left(q^{r\alpha}(t) - C(t)\right)^2\right\rangle_{r\alpha} \tag{6}$$

where $\mathrm{Var}_x[f] = \langle (f - \langle f\rangle_x)^2\rangle_x$ and $\langle\ldots\rangle_x$ is an average over $x$.

$\chi_4(t)$ has a maximum at a time $t \sim \tau$ and $\chi_4(\tau)$ is commonly found to increase continuously as the temperature decreases [24]. This susceptibility has been linked to the growth of a kinetic correlation length [25–27].

In 2007, Berthier and Jack [11] showed that $\chi_4(t)$ could be resolved, by the law of total variance, into two distinct components

$$\chi_4(t) = \chi_{iso}(t) + \chi_{conf}(t) \tag{7}$$

where

$$\chi_{conf}(t) = N\,\mathrm{Var}_r[\langle q^{r\alpha}(t)\rangle_\alpha] = N\,\mathrm{Var}_r[\pi_r(t)] \tag{8}$$

and

$$\chi_{iso}(t) = \langle \chi_{\mathrm{iso}}^r\rangle_r = N\,\langle \mathrm{Var}_\alpha[q^{r\alpha}(t)]\rangle_r, \tag{9}$$

where $\chi_{\mathrm{iso}}^r = N\mathrm{Var}_\alpha[q^{r\alpha}(t)]$ is the isoconfigurational susceptibility associated with a single configuration.

Figure 1 illustrates the successive averaging procedures leading from individual relaxation trajectories to the collective propensity and the standard relaxation function, together with the corresponding decomposition of fluctuations into isoconfigurational and configurational contributions. The two susceptibilities distinguish trajectory-to-trajectory fluctuations at fixed initial configuration, measured by $\chi_{iso}(t)$, from fluctuations of the collective propensity between initial configurations, measured by $\chi_{conf}(t)$. The former quantifies the dynamical variability that remains after fixing the initial structure. If the initial configuration completely determined the collective dynamics, $\chi_{iso}(t)$ would vanish. Conversely, if the initial configuration had no influence on the collective dynamics, $\chi_{conf}(t)$ would vanish.

As the temperature decreases, the relaxation time increases rapidly (Figure 2a), accompanied by a progressive growth of dynamical fluctuations. Figure 2b shows that the three susceptibilities, $\chi_4(t), \chi_{conf}(t)$ and $\chi_{iso}(t)$, develop increasingly pronounced maxima at later times, of the order of the relaxation time $\tau$. Their values at $t = \tau$, shown in Figure 2c, demonstrate that the growth of $\chi_4(t) = \chi_{conf}(t) + \chi_{iso}(t)$ upon cooling is driven by both increasing differences between initial configurations and increasingly strong trajectory-to-trajectory fluctuations at fixed initial configuration. Their relative importance, quantified by the ratio

$$R(T) = \frac{\chi_{iso}(\tau)}{\chi_4(\tau)} = \frac{\chi_{iso}(\tau)}{\chi_{iso}(\tau) + \chi_{conf}(\tau)} \tag{10}$$

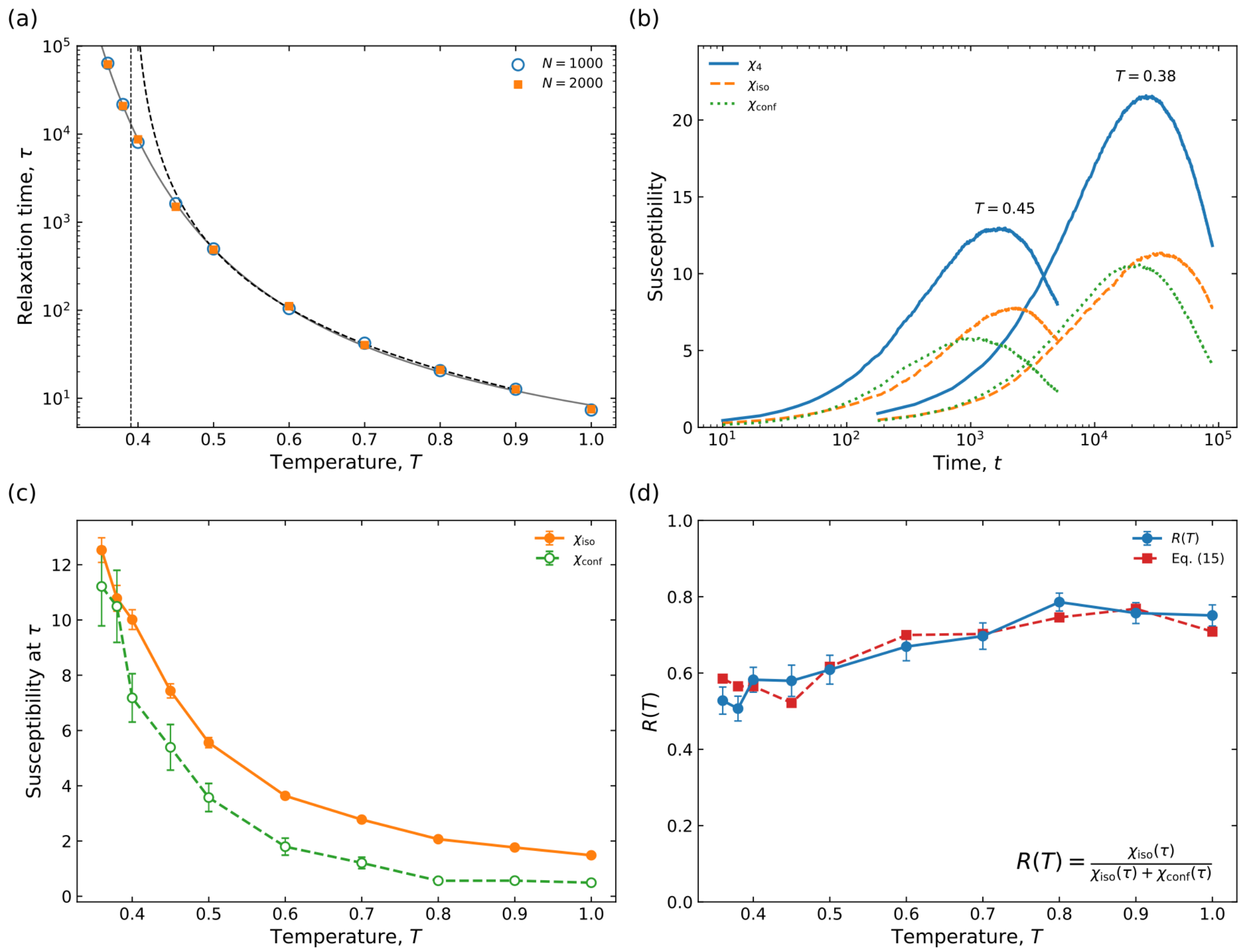

*Figure 2. a) Relaxation time, $\tau$, as a function of the temperature T, for two system sizes. The dashed line is a fit to the mode coupling prediction, $\tau \propto (T - T_c)^{-\gamma}$, from which we estimate $T_c \simeq 0.39, \gamma \simeq 2.4$. The line is a guide to the eye. b) The three variances, $\chi_4(t), \chi_{conf}(t)$ and $\chi_{iso}(t)$ plotted as a function of the time t, at the indicated temperatures. c) $\chi_{conf}(\tau)$ and $\chi_{iso}(\tau)$ as a function of temperature. d) Fraction of the total susceptibility associated with isoconfigurational fluctuations, $R(T) = \frac{\chi_{iso}(\tau)}{\chi_4(\tau)}$, together with the prediction from Eq. 15, using the first-passage-time analysis. Data in (b)-(d) refer to N=1000.*

is shown in Figure 2d. While the configurational contribution becomes relatively more important upon cooling, the ratio tends to level off at $R(T) \simeq 0.5$ at low temperatures, demonstrating that isoconfigurational fluctuations remain a substantial component of the collective dynamical fluctuations even in the deeply supercooled regime. This result also holds in 3D (see Appendix A).

Importantly, these collective susceptibilities reflect both the magnitude and spatial correlations of particle-level fluctuations. We shall distinguish these two contributions by examining the magnitude of microscopic fluctuations in Section 3.2 and their spatial correlations in Section 5.

## 3.2 Microscopic fluctuations

The preceding analysis shows that fluctuations between trajectories starting from the same configuration make a substantial contribution to the collective dynamical susceptibility. Since collective fluctuations reflect both the magnitude and spatial correlations of particle-level fluctuations, we now consider their magnitude at the microscopic level. This allows us to quantify how much of the dynamical variability of individual particles remains after fixing the initial configuration.

Since the particle overlap is binary, the dynamical variability of individual particles, corresponding to the overlap variance over the ensemble of particles, initial configurations, and dynamical realizations is $\mathrm{Var}[o_i^{r\alpha}(t)] = C(t)(1 - C(t))$. Applying the rule of total variance successively over trajectories, particles and initial configurations, the variance is decomposed as

$$C(t)(1 - C(t)) = \langle \mathrm{Var}_\alpha[o_i^{r\alpha}(t)|i, r] \rangle_{i,r} + \langle \mathrm{Var}_\mathrm{i}[\pi_{ri}(t)|r] \rangle_r + \mathrm{Var}_\mathrm{r}[\pi_r(t)] \qquad (11)$$

The three terms quantify, respectively, trajectory-to-trajectory fluctuations at fixed particle and initial configuration, spatial heterogeneity of the microscopic propensity, and configuration-to-configuration fluctuations of the collective propensity. Hence, the first term is a measure of the variation of the dynamics of individual particles between trajectories in each isoconfigurational ensemble, while the last two terms specifically reflect structural influence on dynamics. The last term $\mathrm{Var}_\mathrm{r}[\pi_r(t)] = \chi_{conf}/N$ vanishes in the thermodynamic limit and it is hence expected to be small.

We evaluate these contributions at the bulk relaxation time $\tau$, where $C(\tau) = 1/2$ and the total microscopic variance is fixed at $1/4$. Figure 3a shows that trajectory-to-trajectory fluctuations dominate across the entire temperature range. Although the contributions associated with the initial structure increase upon cooling, together they account for less than 30% of the total variance even at the lowest temperature. Thus, fixing the initial structure leaves most of the microscopic dynamical variability undetermined. This implies that, while the propensity may be predictable from structure, the outcome of an individual trajectory is not, which bounds the accuracy of structure-based predictors of single-trajectory relaxation.

One point of interest, the spatial variation of the microscopic propensity is widely used as evidence of structural influence on dynamics and, yet, the spatial variance in this propensity exhibits on a modest increase on cooling in Figure 3a. We find the variance on its own fails to properly capture the changes in the propensity field as $T$ decreases. In Figure 3b we show that the shape of the propensity distribution changes significantly, from a Gaussian distribution at high $T$ to a broad block-like distribution at low temperatures.

The spatial coarsening of the propensity also increases with cooling, as shown in Figure 3c. We shall quantify this spatial organization through correlation lengths in Section 5.

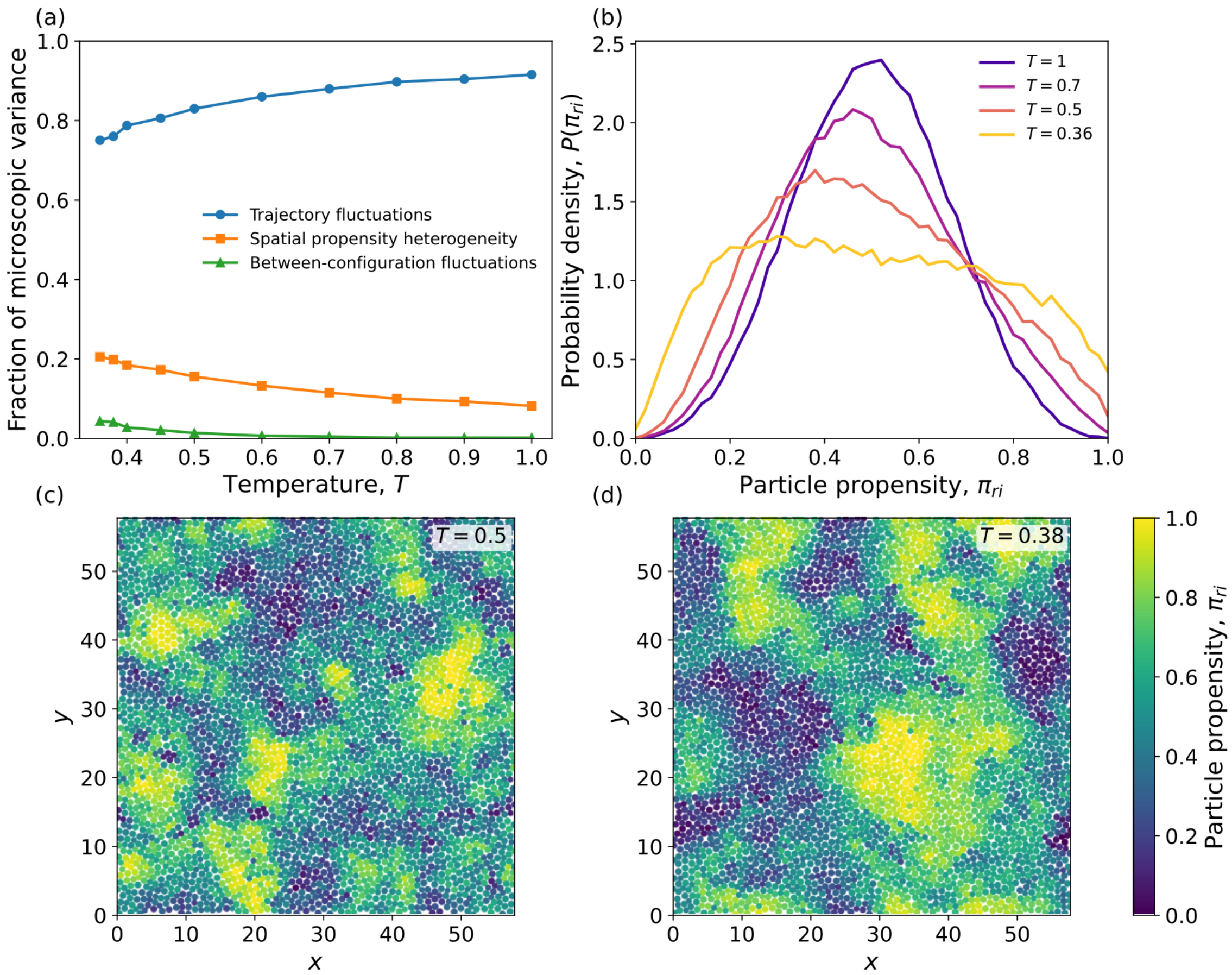

*Figure 3. (a) Decomposition of the microscopic overlap variance into trajectory-to-trajectory fluctuations at fixed initial configuration, spatial heterogeneity of the microscopic propensity, and configuration-to-configuration fluctuations, shown as fractions of the total variance at $C(\tau) = 0.5$. (b) Probability distributions of the microscopic propensity $\pi_{ri}(\tau)$ at selected temperatures. (c,d) Representative spatial maps of the microscopic propensity at $T = 0.5$ and $T = 0.38$, respectively. N=1000 in panels (a) and (b), N=4000 in panels (c) and (d).*

# 4 Isoconfigurational fluctuations follow simple first-passage-time statistics

The preceding section established that trajectory-to-trajectory fluctuations at fixed initial configuration constitute a substantial component of dynamical variability. To investigate their origin, we characterize collective relaxation through the statistics of first-passage times, defined as the times at which individual relaxation trajectories reach a prescribed overlap threshold. First-passage-time statistics are widely used in chemistry and biochemistry [28–30] and have also been employed to investigate particle-level relaxation and persistence in supercooled liquids and glasses [31–33]. Here, we extend this approach to collective relaxation, examining the distributions of first-passage times within individual isoconfigurational ensembles and their dependence on the initial configuration. We then show that these distributions can be described by a simple stochastic model in which relaxation proceeds through a sequence of independently occurring events.

## 4.1 First-passage-time-statistics

For each initial configuration $r$, we define the first-passage time $\tau_1^{r\alpha}$ as the time at which the collective overlap $q^{r\alpha}(t)$ of trajectory $\alpha$ first reaches 1/2. Similarly, we define $\tau_r$ as the time at which the collective propensity $\pi_r(t)$ reaches the same threshold. To characterize the distribution of first-passage times within each isoconfigurational ensemble, we use its median $\tau_{1r}^{(50)}$ and central quantile width $\Delta\tau_{1r} = \tau_{1r}^{(70)} - \tau_{1r}^{(30)}$. The median first-passage time closely agrees with the relaxation time $\tau_r$ of the corresponding collective propensity, as shown in Appendix B. Given the 50 trajectories available for each initial configuration, quantiles provide a robust characterization of the distribution while reducing sensitivity to poorly sampled tails and exceptionally long relaxation times.

Figure 4a shows that, at each temperature, the quantile width increases approximately proportionally to the relaxation time of the corresponding collective propensity, $\Delta\tau_{1r} \simeq a(T)\tau_r$. The notion that structure can influence the mean time scale has been appreciated for over 30 years. Here, for the first time, we see that the choice of initial configuration also influences the degree of dynamical variability observed within an isoconfigurational ensemble. Specifically, it appears that more slowly relaxing configurations tend to exhibit a higher degree of kinetic variability across independent dynamical realizations.

We similarly characterize the distribution of the relaxation time of the collective propensities associated with different initial configurations with its central quantile width $\Delta\tau_r = \tau_r^{(70)} - \tau_r^{(30)}$. Figure 4b compares the temperature dependence of the normalized configurational width, $\Delta\tau_r/\tau$, with the proportionality coefficient $a(T)$, which characterizes the relative isoconfigurational width. Both quantities increase markedly upon cooling, indicating that the variability of relaxation times grows relative to their characteristic timescales, both within individual isoconfigurational ensembles and between different initial configurations. The two measures become comparable at the lowest temperatures investigated.

These observations indicate that the initial configuration primarily sets the characteristic timescale of relaxation, while the relative width of the first-passage-time distribution is approximately configuration-independent at fixed temperature and increases upon cooling.

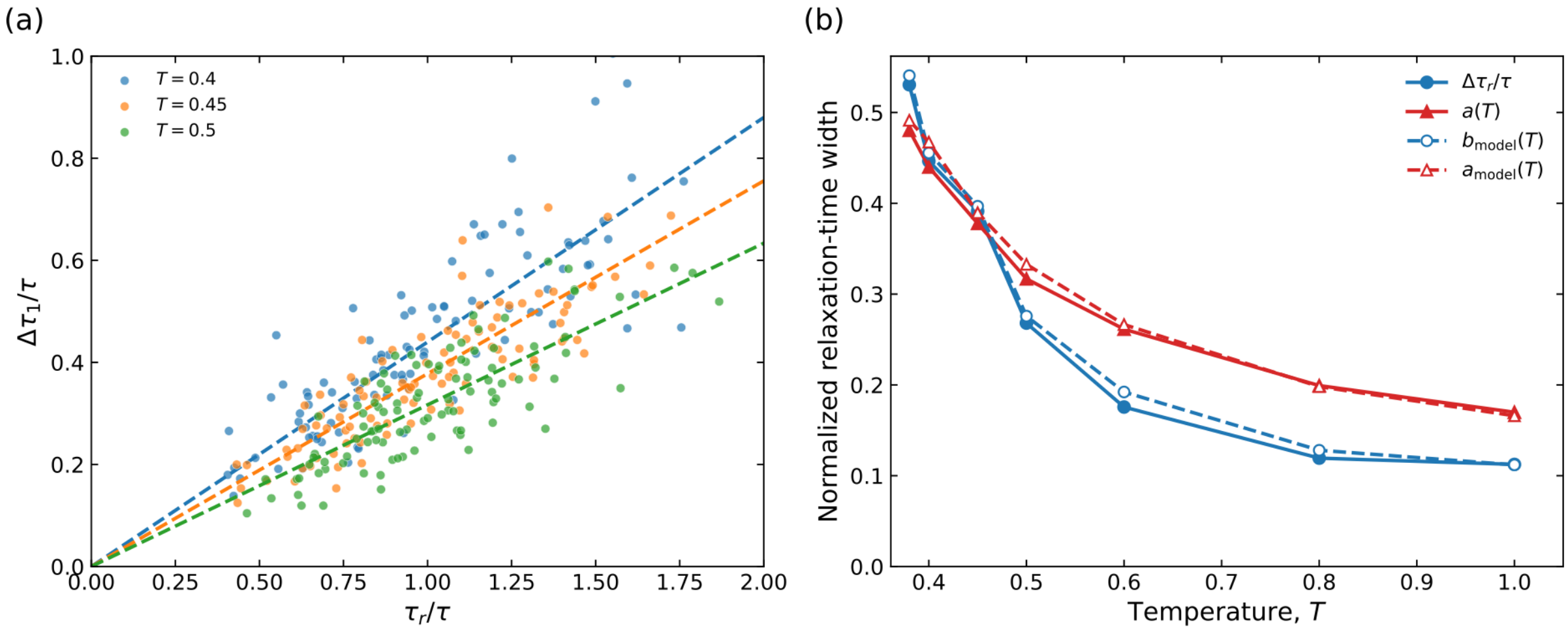

*Figure 4. (a) Central quantile width of the isoconfigurational first-passage-time distribution, $\Delta\tau_1$, plotted against the propensity relaxation time for the same configuration, $\tau_r$, for different temperatures. Each point represents an initial configuration, and both quantities are normalized by the relaxation time $\tau$. Dashed lines are linear fits through the origin, $\Delta\tau_1 = a(T)\tau_r$. (b) Temperature dependence of proportionality coefficient $a(T)$, and of the central quantile width of the distribution of $\tau_r$ normalized by the relaxation time. Dashed lines $a_{model}(T)$ and $b_{model}(T)$,, are predictions from the first-passage time model developed in Section 4.2.*

The first-passage time description of the dynamic fluctuations is directly related to the susceptibilities defined above. Indeed, the fluctuations of the first-passage time $\tau_1^{r\alpha}$ among trajectories generated from the same configuration $r$ are directly related to the iso-configurational fluctuations of the overlap. To leading order,

$$\mathrm{Var}_\alpha[\tau_1^{r\alpha}] \simeq \frac{\chi_{\mathrm{iso}}^r(\tau_r)}{N[\pi_r'(\tau_r)]^2}, \tag{12}$$

where $\pi_r'$(t) denotes the time derivative at time t. The derivation and the approximations involved are given in Appendix C. Similarly, fluctuations of $\tau_r$ among different initial configurations are related to the configurational susceptibility,

$$\mathrm{Var}_\mathrm{r}[\tau_r] \simeq \frac{\chi_{\mathrm{conf}}(\tau)}{N[C'(\tau)]^2}, \tag{13}$$

Thus, $\chi_{\mathrm{iso}}^r$ is related to the trajectory-to-trajectory variability of the relaxation time at fixed initial structure, whereas $\chi_{\mathrm{conf}}$ quantifies the configuration-to-configuration variability of its characteristic value. Combining the two contributions through the law of total variance gives

$$\chi_4(\tau) \simeq N[C'(\tau)]^2 \mathrm{Var}_{\mathrm{r},\alpha}[\tau_1], \tag{14}$$

a relation that makes explicit that the standard deviation of the first-passage-time distribution decreases as $N^{-1/2}$. This relation connects the first-passage-time description directly to the decomposition of the collective dynamical fluctuations introduced in Section 3. We can eliminate the dependence on $C'(\tau)$ by calculating the ratio $R(T)$,

$$R(T) = \frac{\chi_{\mathrm{iso}}(\tau)}{\chi_4(\tau)} = \frac{\langle \mathrm{Var}_\alpha[\tau_1|r]\rangle_r}{\mathrm{Var}_{\mathrm{r},\alpha}[\tau_1]} \simeq \frac{\langle\Delta\tau_1^2\rangle_r}{\langle\Delta\tau_1^2\rangle_r + \Delta\tau_r^2}, \tag{15}$$

where in the last step, we have approximated the variances with the squared central quantile widths. The calculation of $R(T)$ using Eq. 15 is shown to reproduce the values obtained directly from ratio of susceptibilities as shown in Figure 2d.

## 4.2 Stochastic relaxation-event model

Our analysis of the relaxation dynamics has revealed two important features. First, the characteristic relaxation time depends on the initial configuration, distinguishing relatively fast and slow configurations at the same temperature. Second, trajectories starting from the same configuration exhibit substantial variability in their first-passage times, with a distribution whose width is approximately proportional to its characteristic relaxation time. Here we show that these observations can be described by a simple stochastic model in which relaxation proceeds through a sequence of effective, independently occurring events. The initial configuration determines the rate of these events, while their stochastic occurrence generates the variability between trajectories. We highlight that the events are assumed to be independent.

A useful starting point is a fixed-rate Poisson process. We assume that relaxation to the threshold requires the accumulation of $m(T,N)$ effective events, occurring independently at a constant rate $\lambda_r$ determined by the initial configuration $r$. The first-passage time $\tau_1^{r\alpha}$ of an individual trajectory is then a random variable corresponding to the waiting time for the $m$-th event. Its probability density is,

$$P(\tau_1^{r\alpha}|\lambda_r, m) = \frac{\lambda_r^m (\tau_1^{r\alpha})^{m-1} e^{-\lambda_r \tau_1^{r\alpha}}}{\Gamma(m)} \tag{16}$$

Eq. 16 is an Erlang distribution [34] with mean $\langle \tau_1^{r\alpha} \rangle_r = \frac{m}{\lambda_r}$ and variance $\mathrm{Var}_r(\tau_1^{r\alpha}) = \langle \tau_1^{r\alpha} \rangle_r^2 / m$. At fixed $m(T,N)$, changing $\lambda_r$ simply rescales time. Consequently, the approximately proportional relationship between the distribution width and its median, observed in Figure 4a, follows directly from the model.

For each temperature and system size, we estimate the parameters $m, \{\lambda_r\}$ by maximizing the joint likelihood as described in Appendix D. Figure 5a shows that $m/N$ decreases markedly on cooling and is approximately independent of system size, confirming the expected extensive scaling of the effective number of events required for the system to relax. The quality of the stochastic description is illustrated in Figure 5b-d, where we compare the empirical survival probabilities with the corresponding Erlang predictions, $S_r(t) = \int_t^\infty P_r(\tau_1)\, d\tau_1$. As a consistency check, Figure 4 shows that the fitted model reproduces the measured relation between the width and median of the first passage time distribution (panel a), and the width of the distribution of the configuration dependent rates (panel b). We note that other waiting-time distributions with a width proportional to the mean, such as the log-normal, would describe the data equally well, and discriminating among them would require comparing relations involving higher moments, which demands a much larger number of trajectories than is available here. For this reason, we consider $m = \langle \tau_1^{r\alpha} \rangle_r^2 / \mathrm{Var}_r(\tau_1^{r\alpha})$ an effective number of events.

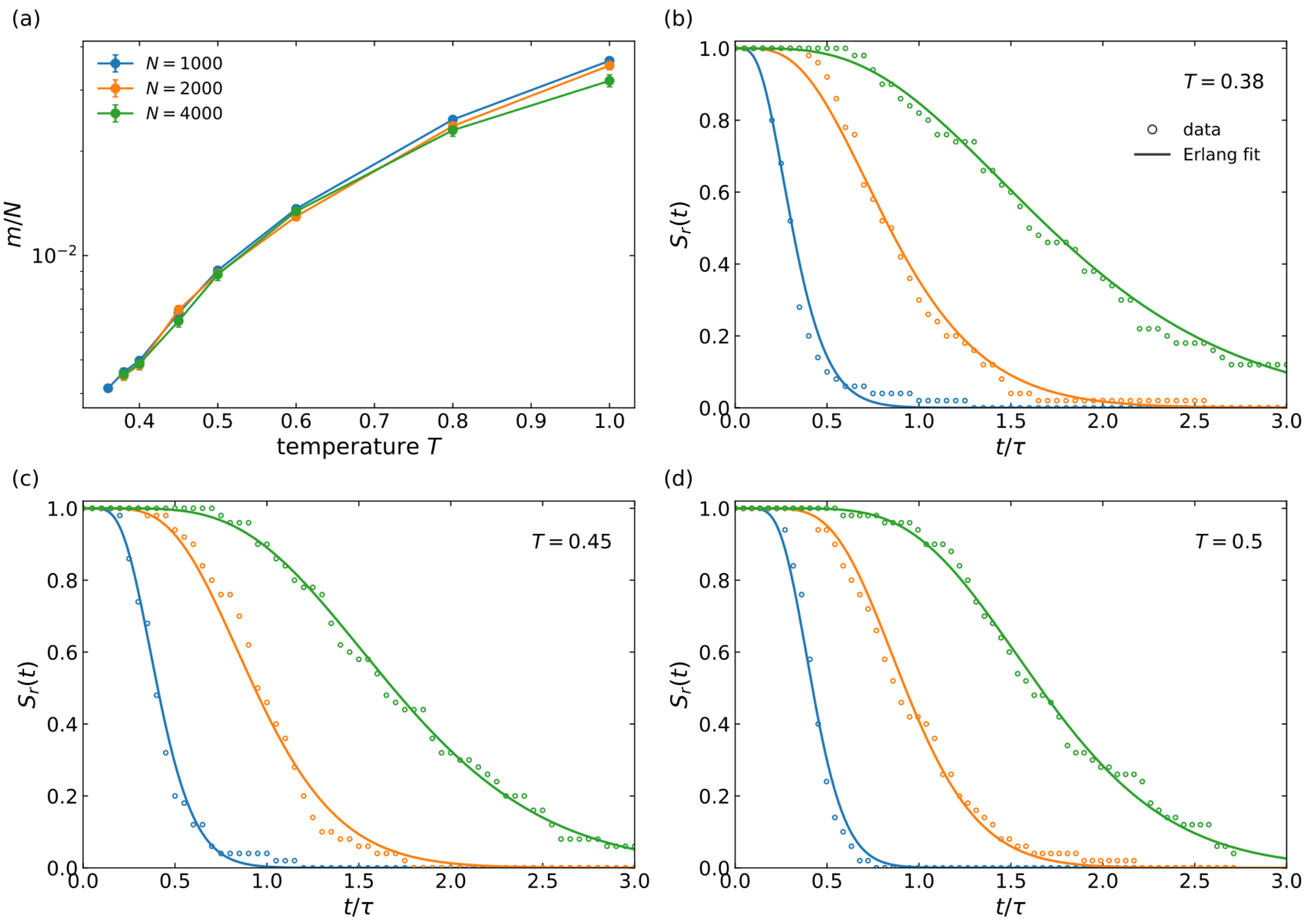

*Figure 5. (a) Scaled effective event number m/N as a function of temperature for three system sizes. (b-d) Empirical first-passage-time survival probabilities for three representative initial configurations at T=0.38, 0.45, and 0.5, compared with the corresponding Erlang model predictions, for N = 1000.*

The model, therefore, describes both sources of relaxation-time variability through two complementary ingredients: the characteristic rates $\lambda_r$, which distinguish different initial configurations, and stochastic event accumulation, which generates fluctuations between trajectories starting from the same configuration. The decrease of $m$ upon cooling, together with its extensive system-size dependence, suggests that each effective relaxation event involves a number of particles $N/m$ increasing from 25 to 250 in the investigated temperature range. These events should not be identified with elementary localized neighbour exchanges, but rather with larger-scale episodes of collective relaxation, potentially comprising multiple elementary rearrangements.

Assuming these events are spatially compact, we associate them with an effective length scale $\xi_m(T) \propto \left[\frac{N}{m(T,N)}\right]^{1/2}$ in two dimensions. In the following section, we compare this length scale with those obtained independently from the spatial correlations of dynamical fluctuations.

# 5 Isoconfigurational fluctuations are spatially correlated

Having established that isoconfigurational fluctuations dominate the microscopic dynamical variability and can be described through a stochastic relaxation-event model, we now investigate their spatial organization. Here, we ask whether fluctuations of different particles about their respective propensities are spatially correlated, and whether the characteristic length scale of these correlations is related to the effective size of the relaxation events inferred from the 1st passage time statistics.

To distinguish correlations associated with the initial structure from those arising within individual dynamical realizations, we introduce the configurational and isoconfigurational fluctuating fields,

$$\delta\pi_{ri} = \pi_{ri} - C, \qquad \delta o_i^{r\alpha} = o_i^{r\alpha} - \pi_{ri} \tag{17}$$

where $C(t)$ is defined in Eq. 5.

The field $\delta\pi_{ri}$ is the fluctuation of the usual propensity field about its ensemble average $C$(t) and therefore probes the spatial correlations associated with the influence of the structure on the dynamics. The second field $\delta o_i^{r\alpha}$ describes the residual trajectory-to-trajectory fluctuations at fixed initial structure; the associated correlation length measures the spatial extent over which particles tend to fluctuate together around their local propensities. The spatial integrals of the covariances associated with these two fields give $\chi_{conf}$ and $\chi_{iso}$, as in Eqs. 8 and 9.

To quantify the spatial extent of these fluctuations, we calculate the structure factors associated with the two fields, $S_{conf}(q)$ and $S_{iso}(q)$. For the configurational contribution, we average over initial configurations, whereas for the isoconfigurational contribution, we average over trajectories and initial configurations. We extract the corresponding correlation lengths $\xi_{conf}(T)$ and $\xi_{iso}(T)$ by fitting the small-wavevector behavior of the structure factors to the Ornstein–Zernike form. All correlation lengths are evaluated at the bulk relaxation time $\mathrm{t} = \tau$. Details of the fitting procedure and representative structure factors are provided in Appendix E.

Fig. 6 shows the temperature dependence of $\xi_{conf}(T)$, which characterizes the spatial correlations of the propensity field, and of $\xi_{iso}(T)$, which measures the spatial extent of correlated trajectory-to-trajectory fluctuations. Both correlation lengths increase upon cooling, demonstrating that the spatial organization of both configurational and trajectory-dependent fluctuations becomes more pronounced as relaxation slows down. However, $\xi_{conf}(T)$ remains systematically larger than $\xi_{iso}(T)$, indicating that the influence of the initial structure extends over larger distances than the correlations among fluctuations within the isoconfigurational ensemble.

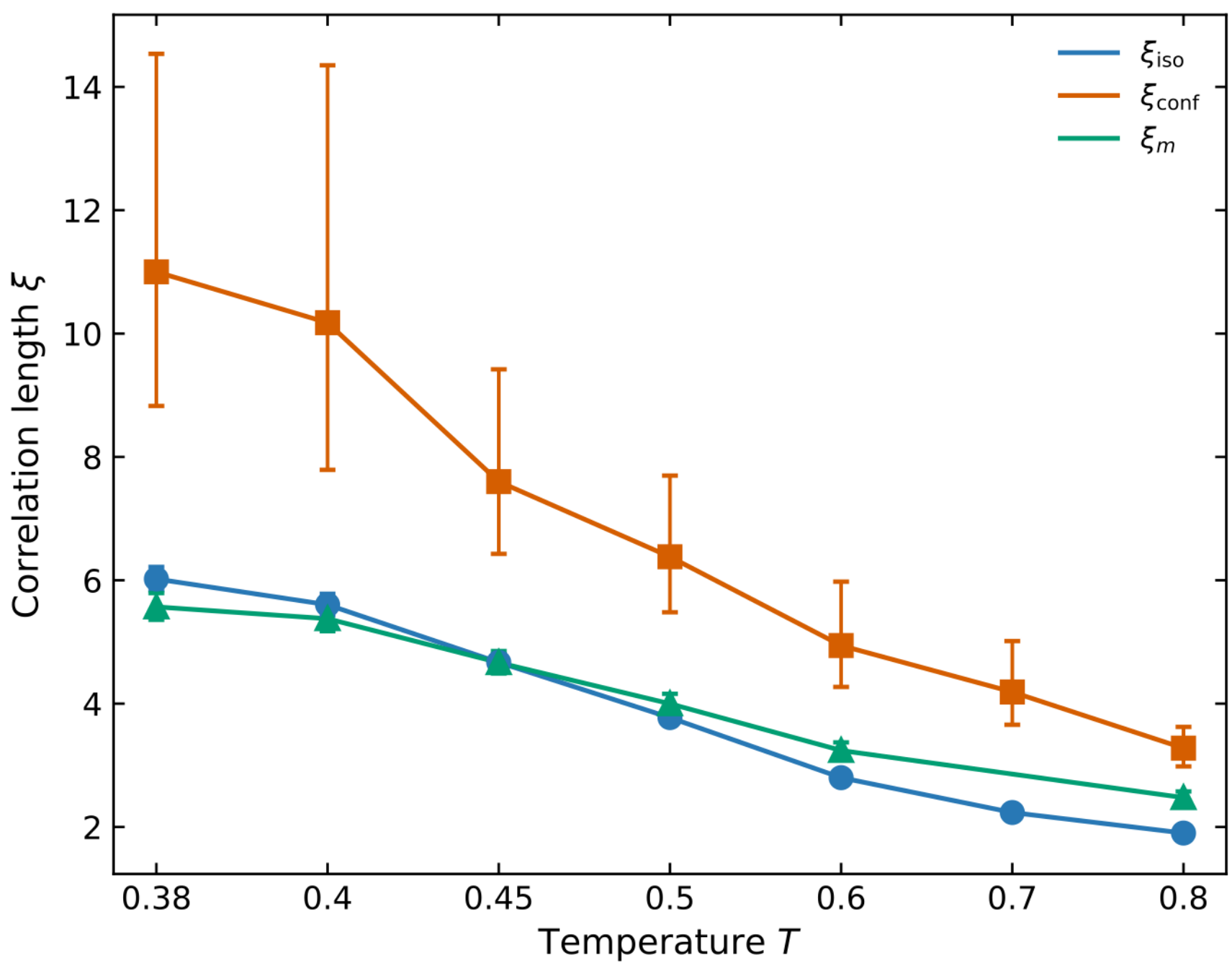

*Figure 6. Temperature dependence of the correlation lengths $\xi_{iso}(T)$, $\xi_{conf}(T)$ and $\xi_m(T)$, with the latter rescaled by $A \simeq 0.42$.*

The figure also compares these lengths with $\xi_m(T) = \rho^{-\frac{1}{2}}\sqrt{\frac{N}{m(T)}}$, where $m(T)$ is the effective number of relaxation events inferred from the first-passage time distributions, as discussed in Sec. 4. Remarkably, $A\xi_m$, with $A \simeq 0.42$ a single temperature-independent proportionality factor, closely follows $\xi_{iso}$ over the investigated temperature range. This agreement connects the temporal statistics of collective relaxation to the spatial correlations of microscopic dynamical fluctuations. It suggests that the effective events underlying the stochastic description of relaxation may be associated with the characteristic spatial scale of isoconfigurational fluctuations, distinct from the larger scale characterizing the structural influence on dynamics.

Figure 6 presents us with two interesting puzzles. The first is that $\xi_{iso} \neq \xi_{conf}$. To understand this point we have plotted in Figure 7 spatial maps at two temperatures of the fluctuations of the propensity $\delta\pi_i$ for a given initial configuration and the fluctuations of the overlap $\delta o_i$ for a single trajectory for the same initial configuration. We find the fluctuations in the overlap exhibit a weak correlation with the propensity – i.e. high propensity regions correspond to weak positive (pink) fluctuations in overlap while low propensity regions see weak negative fluctuations (pale blue). These correlations, however, are not the dominant features of the overlap fluctuations. The large positive and negative fluctuations take the form of moderately compact domains located at the boundaries between high and low propensity regions. Some of these fluctuations have been circled in Figure 7. The length $\xi_{iso}$ describes these domains. The increase in $\xi_{iso}$ with decreasing T appears to be related to the increasing coarse graining of the propensity distribution and the associated growth in the length of the borders between high and low propensity regions.

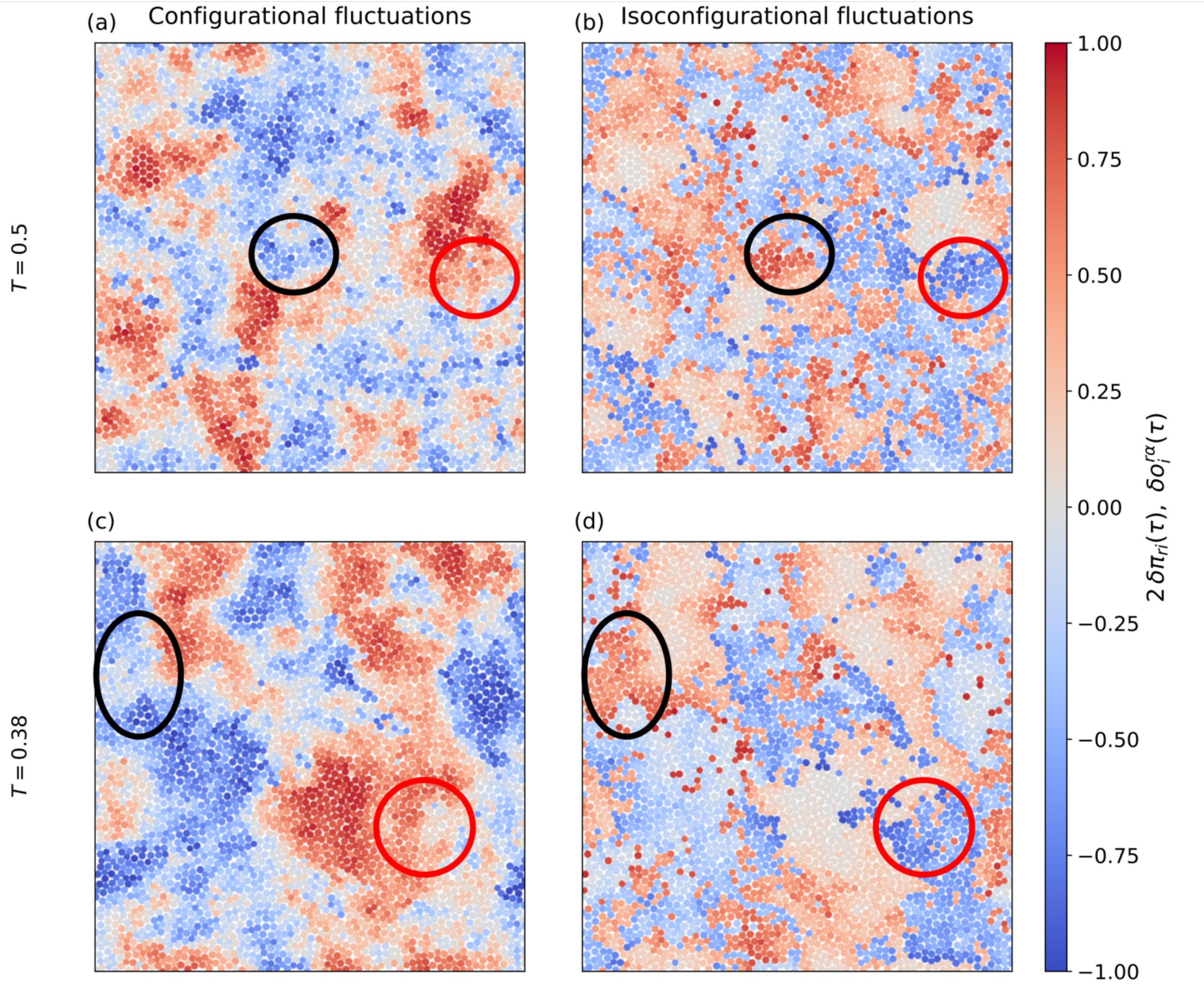

*Figure 7. Representative maps of (a,c) the configurational fluctuation field, $\delta\pi_{ri}(\tau) = \pi_{ri}(\tau) - C$, scaled by a factor of 2 as |δπ_ri (τ)|<1/2 and (b,d) the isoconfigurational fluctuation field, $\delta o_i^{r\alpha}(\tau) = o_i^{r\alpha}(\tau) - \pi_{ri}(\tau)$, at T=0.5 (a,b) and T=0.38 (c,d) , for N = 4000. At each temperature, the two maps correspond to the same initial configuration, with the isoconfigurational field evaluated for a single trajectory $\alpha$. Colours indicate the magnitude and sign of the fluctuations.*

This brings us to our second puzzle from Figure 6, namely the result that $\xi_{iso} \simeq \xi_m$, at least in terms of the temperature dependence. This rough equality provides some support for the interpretation of the 'events' referred to in the stochastic model of the 1st passage time distribution, i.e. Eq. 16, as being real. If the events are real then it seems natural to associate the compact deviations of the overlap from the propensity, shown in Figure 7, as spatial representations of the stochastic events themselves. Whether or not, this simple identification stands up to further examination, the fact that the iso-fluctuations provide us with a new characteristic length, one explicitly connected to how things move as opposed to where in space the movements take place, is an exciting development.

# 6 Conclusions

In this paper we have focused on the variation in dynamics between different trajectories from a common starting configuration. In summary, our findings are that these iso-fluctuations are a) large, b) reasonably described a simple stochastic sequence of events, and c) characterized by a new length scale related to compact fluctuations in dynamics that occur in the boundaries between regions between high and low propensity.

There is, in the literature, a focus on the propensity as a means of analyzing cooperative dynamics in supercooled liquids that tends to reinforce a perception that the cooperative dynamics in a supercooled liquid is essential determined by the underlying structure. The fact that the actual trajectories by which a system relaxes exhibit significant fluctuations about the propensity challenges this notion of structural determinism, a point made previously [11]. Structure exerts its significant influence on the average over trajectories, i.e. the propensity, by spatially locating the bounds on the propensity, those particles with particularly low or high propensity for motion. Regions of intermediate propensity correspond to regions where the structure exerts little to no constraint on the dynamics.

The variability of the isoconfigurational fluctuations is nicely captured by the 1st passage time distribution that we have introduced. We find that the width of the first-passage-time distribution for trajectories originating from the same configuration is approximately proportional to their characteristic relaxation time, with a proportionality coefficient that increases upon cooling. These distributions are well described by a stochastic model in which relaxation requires the accumulation of a number of uncorrelated events at a configuration-dependent rate. The resulting Erlang distribution reproduces the first-passage-time statistics. The decrease of the number of events per particle upon cooling suggests that the effective size of independent relaxation events increases as the liquid becomes more deeply supercooled. The premise on which this stochastic model is that the choice of the initial momenta represents a random choice of a particular random sequence of dynamic events. The mechanism of this selection process remains an open question.

These conclusions bring us to the question of the physical characteristics of the isoconfigurational fluctuations. We find that they exhibit a correlation length that grows slower, on cooling, than that associated with the propensity (and, hence, the structural influence). Remarkably, this length scale agrees, up to a temperature-independent proportionality factor, with the effective length inferred independently from the stochastic first-passage-time model. We have shown that these fluctuations take the form of compact domains located in regions of intermediate propensity. The growth in $\xi_{iso}$ with decreasing T appears to be associated with some combination of the size of these domains and the changing topology of the regions of intermediate propensity. Unravelling this subtle interplay of stochastic events and structural influence represents a fundamental project for future work.

**Acknowledgements**

M.P.C. acknowledges supported by the Singapore Ministry of Education through grants

MOE-T2EP50224-0016, RG152/23, and RG153/24.

# Appendix A. Three-dimensional results

We have investigated the decomposition of the susceptibility $\chi_4(t) = \chi_{iso}(t) + \chi_{conf}(t)$ for the standard Kob-Andersen binary mixture in three dimensions.

We focused on two set of data.

1) Configurations of $N = 4096$ particles. For $T \geq 0.44$ these are the configurations of Ref. [35]. For $T = 0.404, 0.408, 0.412, 0.416$, and $T = 0.420$ we prepared equilibrium configurations via parallel tempering, checking equilibration as in Ref. [36]. At each temperature, we consider 200 configurations, and for each configuration, an iso-configurational ensemble of 8 trajectories.
2) Configurations of N = 4000 particles, equilibrated to low temperatures, kindly shared by the authors of Ref. [23]. At each temperature we consider 30 independent configurations and, for each configuration, an iso-configurational ensemble of 50 trajectories. The interaction potential uses a quadratic rather than a linear smoothing at the cutoff.

We probe relaxation via the self-scattering function $F_s(q, t)$ at the wave-vector of the first peak of the structure factor [11], and define the relaxation time as $F_s(q, \tau) = 1/e$.

Figure 8 shows that $R(T) = \chi_{iso}(\tau)/\chi_4(\tau)$ decreases on cooling, reaches a minimum at around $T \simeq 0.42$, and then increases slightly at lower temperatures. This nonmonotonic behavior is consistent with the saturation observed in the two-dimensional model, indicating that the relative contribution of isoconfigurational fluctuations remains substantial in the deeply supercooled regime. In the figure we also report the earlier high-temperature ($T \geq 0.45$) results by [11]. In the overlapping temperature range the two datasets show the same trend, with our $R$ values somewhat smaller. Possible origins of the quantitative difference are the system size (Ref [11] used N = 1000) and the statistics used to estimate the susceptibilities.

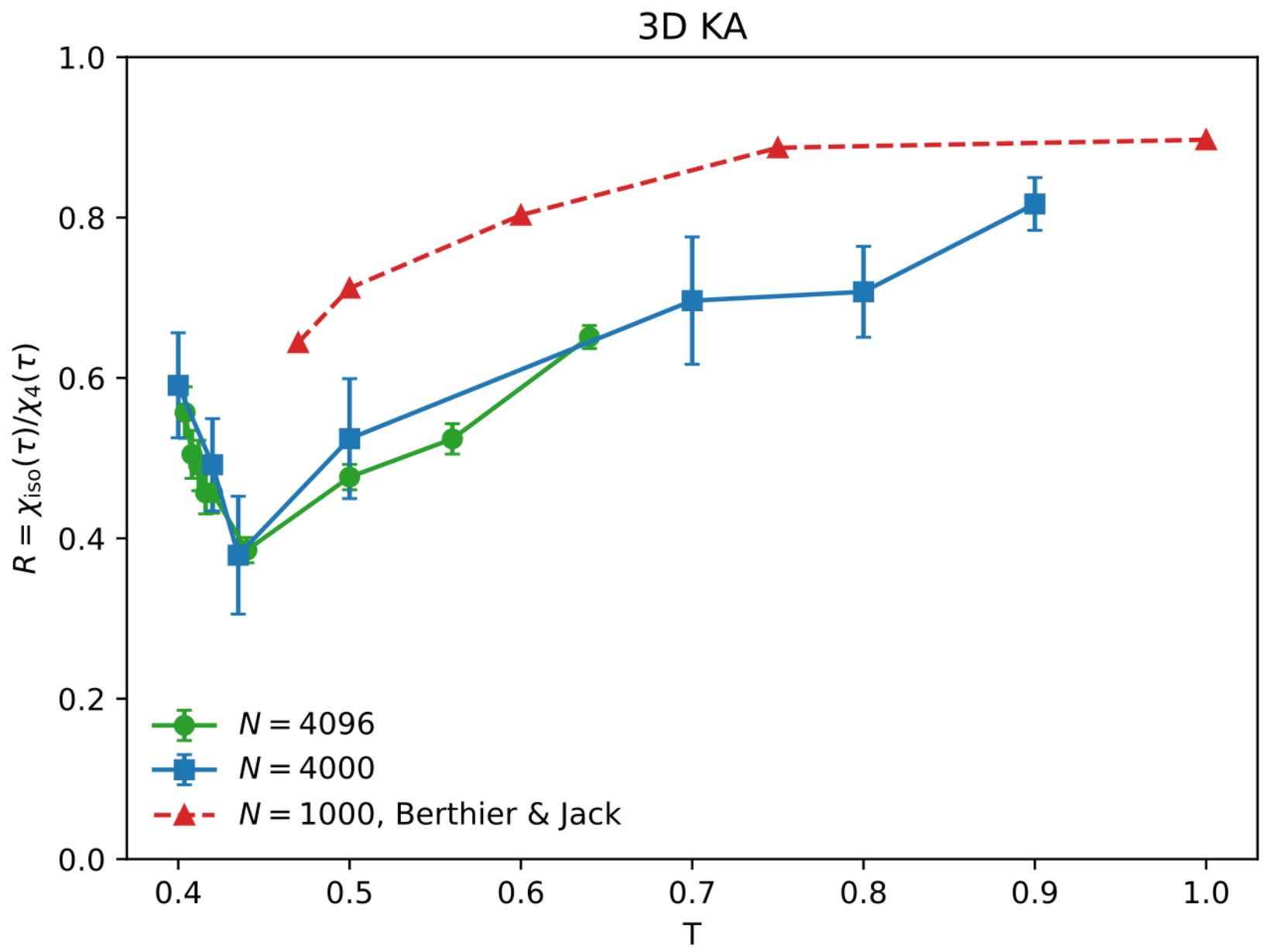

*Figure 8. Temperature dependence of* $R(T) = \frac{\chi_{iso}(T)}{\chi_4(T)}$ *for the 3d KA mixture, whose estimated mode-coupling critical temperature is,* $T_c$*=0.435. We also illustrate the data from Fig. 2 of Berthier and Jack [11] obtained for a system with N = 1000 particles (statistics not specified) at* $T > T_c$*.*

# Appendix B. Median 1st passage time

The median first-passage time $\tau_{1r}^{(5)}$ coincides with the relaxation time $\tau_r$ of the collective propensity, defined by $\pi_r(\tau_r) = 1/2$. To leading order, the first-passage time of a trajectory deviates from $\tau_r$ by $\delta q^{r\alpha}(\tau_r)/|\pi_r'(\tau_r)|$, and $\delta q_r^{\alpha}$ has zero mean, so the distribution is centered on $\tau_r$. We use the median because it is robust to the rare, very long first-passage times and to right-censored trajectories. Figure 9 confirms the agreement: each point, one per initial configuration, falls on the identity line over several orders of magnitude in relaxation time.

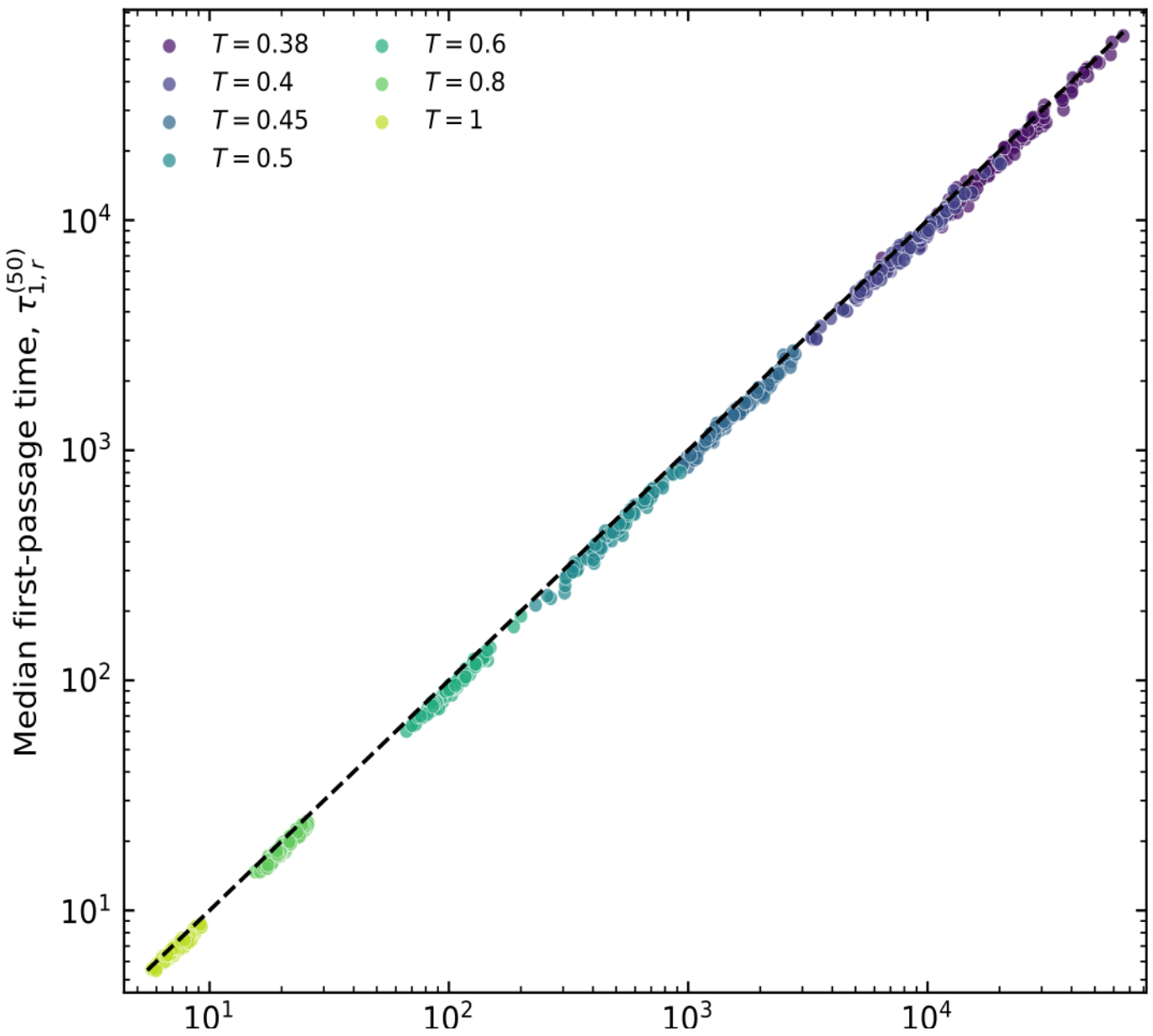

*Figure 9. Median first-passage time $\tau_{1,r}^{(50)}$ versus the relaxation time $\tau_r$ of the corresponding collective propensity, for different temperatures. Each point represents an initial configuration. The dashed line indicates equality, $\tau_{1,r}^{(50)} = \tau_r$, highlighting the close agreement between the two timescales over several orders of magnitude.*

# Appendix C. Susceptibilities and 1st passage time

Here we derive the relation between fluctuations of the overlap and those of the first-passage time. For a fixed initial configuration $r$, we consider the configurational propensity$\pi_r(t) = \langle q^{r\alpha}(t)\rangle_\alpha$, which satisfies $\pi_r(\tau_r) = \frac{1}{2}$.

An individual trajectory as an overlap $q^{r\alpha}(t)$ and a first-passage-time $\tau_1^{r\alpha}$ which deviates, respectively, from the iso-configurational averaged overlap $\pi_r(t)$ and its crossing time $\tau_r$. That is,

$$q^{r\alpha}(t) = \pi_r(t) + \delta q^{r\alpha}(t); \qquad \tau_1^{r\alpha} = \tau_r + \delta\tau_1^{t\alpha}(r,p) \tag{A1}$$

By definition of first-passage time, $q^{r\alpha}(\tau_1^{r\alpha}) = 1/2$. Therefore,

$$q_{r,p}(t_r + \delta\tau_1) = \pi_r(t_r) + \pi_r'(t_r)\delta\tau_1 + \delta q_{r,p}(t_r) + \partial_t \delta q_{r,p}(t_r)\delta\tau_1 = \frac{1}{2} \tag{A2}$$

Since $\pi_r(\tau_r) = 1/2$, and since both $\delta q$ and $\delta\tau_1$ scale as $N^{-1/2}$, the last term is of second order in the fluctuations and may be neglected. We therefore obtain

$$\delta\tau_1^{r\alpha} \simeq -\frac{\delta q^{r\alpha}(\tau_r)}{\pi_r'(\tau_r)} \tag{A3}$$

Because the relaxation function decreases with time, $\pi_r'(\tau_r) < 0$. Thus, trajectories for which $q^{r\alpha}(\tau_r) > \pi_r(\tau_r)$ have $\delta\tau_1^{r\alpha} > 0$: trajectories that have relaxed less than average at time $\tau_r$ cross the threshold later.

Averaging the above relation over trajectories gives $\langle\delta\tau_1^{r\alpha}\rangle_\alpha = 0$ because $\langle\delta q^{r\alpha}(\tau_r)\rangle_\alpha = 0$, so that to this order $\langle\tau_1^{r\alpha}\rangle \simeq \tau_r$. Taking the variance over trajectories at fixed initial configuration $r$, we recover Eq. 12 in the main text,

$$\mathrm{Var}_\alpha[\tau_1^{r\alpha}] \simeq \frac{\mathrm{Var}_\alpha[q^{r\alpha}(t)]}{[\pi_r'(\tau_r)]^2} \simeq \frac{\chi_{\mathrm{iso}}^r(\tau_r)}{N[\pi_r'(\tau_r)]^2} \tag{A4}$$

With $\chi_{\mathrm{iso}}^r(t) = N\,\mathrm{Var}_\alpha[q^{r\alpha}(t)]$ the isoconfigurational susceptibility for structure $r$. Thus, if $\chi_{\mathrm{iso}}^r(\tau_r)$ remains finite for $N \to \infty$, $\mathrm{Var}_\alpha[q^{r\alpha}(t)] \propto N^{-1}$, and the standard deviation of the first-passage-time distribution at fixed configuration decreases as $N^{-1/2}$.

We now consider fluctuations between different initial configurations. For each configuration, we write

$$\pi_r(t) = C(t) + \delta\pi_r(t) \tag{A5}$$

where $C(t) = \langle\pi_r(t)\rangle_\mathrm{r}$ is the standard correlation function, and $\mathrm{C}(\tau) = 1/2$.

Since $\pi_r(\tau_r) = 1/2$, we expand around the bulk relaxation time:

$$\pi_r(t_r) = \mathrm{C}(\tau) + \mathrm{C}'(\tau)(t_r - \tau) + \delta\pi_r(\tau) \tag{A6}$$

Obtaining

$$t_r - \tau = -\frac{\delta\pi_r(\tau)}{C'(\tau)} \tag{A7}$$

Taking the variance over initial configurations gives Eq. 13 in the main text,

$$\mathrm{Var_r}[\tau_r] \simeq \frac{\mathrm{Var_r}[\pi_r(\tau)]}{[C'(\tau)]^2} = \frac{\chi_{\mathrm{conf}}(\mathrm{t})}{N[C'(\tau)]^2}$$

with $\chi_{conf}(\mathrm{t}) = N\ \mathrm{Var_r}[\pi_r(\tau)]$ the configurational susceptibility. Thus, provided that $\chi_{conf}(\mathrm{t})$ remains finite in the thermodynamic limit, $\mathrm{Var_r}[\tau_r] \propto N^{-1}$.

The total first-passage-time fluctuations contain both contributions. By the law of total variance

$$\mathrm{Var_{r,\alpha}}[\tau_1^{r\alpha}] = \langle \mathrm{Var_\alpha}[\tau_1^{r\alpha}]\rangle_\mathrm{r} + \mathrm{Var_r}[\langle \tau_1^{r\alpha}\rangle_\alpha]$$

In the large-N limit, $\langle \tau_1^{r\alpha}\rangle_\alpha \simeq \tau_r$, and the configuration-to-configuration variations of the local relaxation slope are small, $\pi'(\tau_r) \simeq C'(\tau)$. Under these approximations

$$\chi_{\mathrm{iso}}(\tau) = \mathrm{N}[C'(\tau)]^2\langle \mathrm{Var}_\alpha[\tau_1^{r\alpha}]\rangle_\mathrm{r} \tag{A8}$$

$$\chi_{\mathrm{conf}}(\tau) = \mathrm{N}[C'(\tau)]^2\mathrm{Var_r}[\tau_r] \tag{A9}$$

and

$$\chi_4(\tau) = \chi_{\mathrm{iso}}(\tau) + \chi_{\mathrm{conf}}(\tau) = \mathrm{N}[C'(\tau)]^2\mathrm{Var_{r,\alpha}}[\tau_1^{r\alpha}] \tag{A10}$$

# Appendix D. Likelihood estimation of the stochastic relaxation-event model

We describe here the procedure used to estimate the parameters of the stochastic relaxation-event model introduced in Sec. 4.2. At each temperature and system size, we assume that the first-passage times $\tau_1^{r\alpha}$ of trajectories originating from configuration $r$ follow an Erlang distribution with shape parameter $m(N,T)$ and configuration-dependent event rate $\lambda_r$,

$$P(\tau_1^{r\alpha}|\lambda,m) = \frac{\lambda^m (\tau_1^{r\alpha})^{m-1} e^{-\lambda\tau_1}}{\Gamma(m)} \qquad \text{(A11)}$$

The parameter $m$, representing the effective number of relaxation events, is common to all initial configurations, whereas $\lambda_r$ accounts for their different characteristic relaxation times.=

For each initial configuration $r$, we consider $n_r$ independently generated trajectories. Assuming that their first-passage times are independent realizations, the probability density of observing a particular set of first-passage times is given by the product of their individual probability densities. Since the initial configurations are also independently sampled, the joint likelihood of the complete dataset is obtained by multiplying these contributions over all configurations,

$$\mathcal{L}(m,\{\lambda_r\}) = \prod_r \prod_{\alpha=1}^{n_r} P(\tau_1^{r\alpha}|m,\lambda_r) \qquad (A12)$$

Maximizing the joint likelihood identifies the parameters under which the observed dataset is most likely.

In a few cases, particularly for small system sizes where the first-passage-time fluctuations are relatively large, some trajectories do not reach the prescribed overlap threshold within the simulation time. We denote by $n_r^c$ the number of such trajectories among the $n_r$ trajectories originating from configuration $r$. Although their first-passage times are not observed, we know that they exceed the maximum simulation time $t_{\max}^{r\alpha}$. These trajectories are therefore right-censored, and their contribution to the likelihood is given by the survival probability

$$S(t_{\max}^{r\alpha}|m,\lambda_r) = \int_{t_{\max}^{r\alpha}}^{\infty} P(t|m,\lambda_r)\, dt = \frac{\Gamma(m,\lambda_r t_{\max}^{r\alpha})}{\Gamma(m)} \qquad (A13)$$

where $\Gamma(m,x)$ is the upper incomplete gamma function. The joint likelihood, accounting for both censored and uncensored trajectories, becomes

$$\mathcal{L}(m,\{\lambda_r\}) = \prod_r \left[\prod_{\alpha\in O_r} P(\tau_1^{r\alpha}|m,\lambda_r)\right]\left[\prod_{\alpha\in C_r} S(t_{\max}^{r\alpha}|m,\lambda_r)\right] \qquad (A14)$$

where $O_r$ and $C_r$ are the set of uncensored and censored trajectories, containing respectively $n_r - n_r^c$ and $n_r^c$ elements.

For a fixed value of $m$, which we allow to vary continuously using the gamma-function extension of the Erlang distribution, we maximize the logarithm of the likelihood with respect to the configuration-dependent rates $\{\lambda_r\}$, evaluating the log likelihood profile

$$\log \mathcal{L}(m) = \max_{\{\lambda_r\}} \log \mathcal{L}(m, \{\lambda_r\}) \qquad (A15)$$

shown in Figure 10. This profile has a well-defined maximum, from which we determine the effective event number $m$ and the associated configuration-dependent rates $\{\lambda_r\}$.

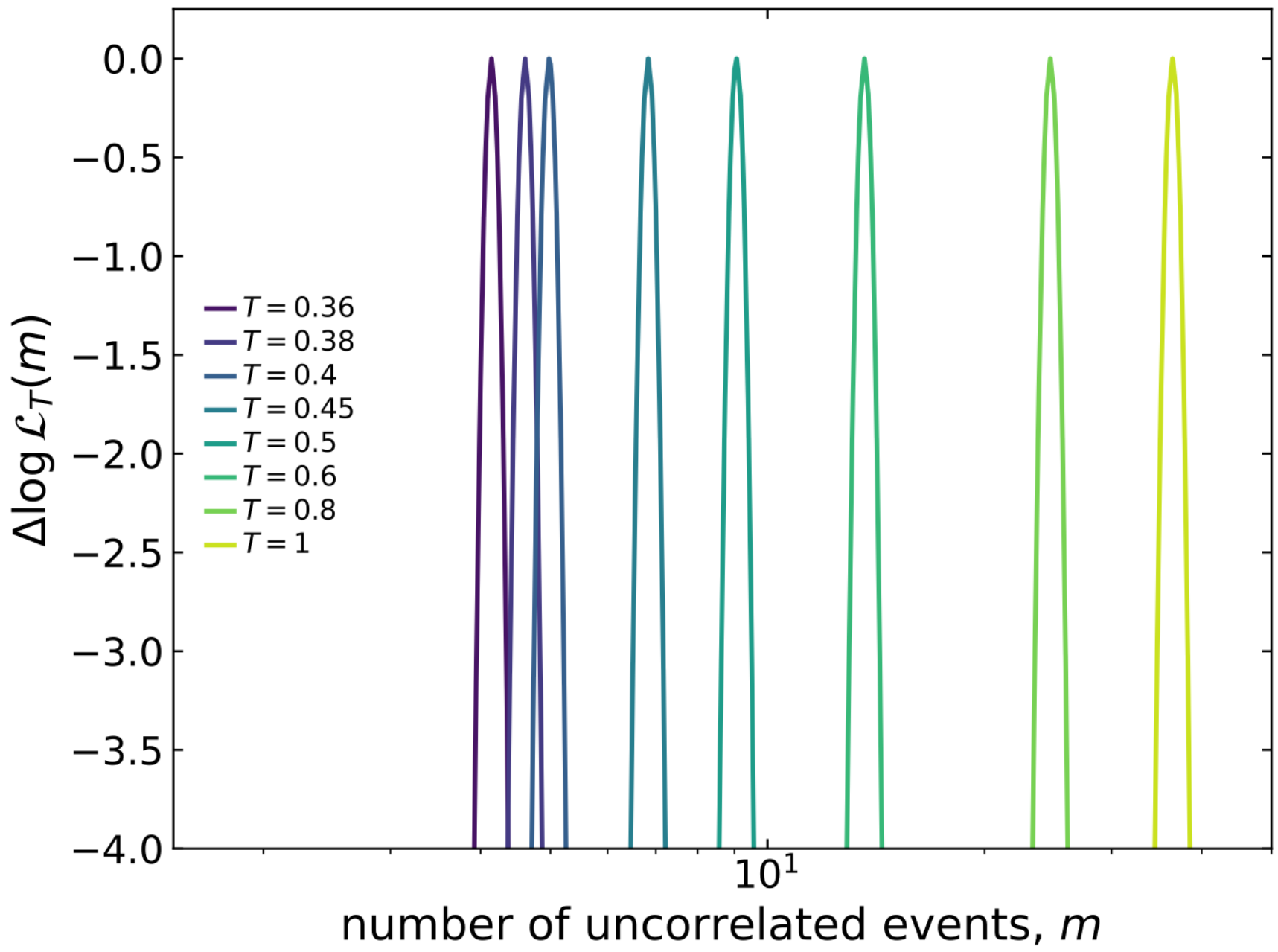

*Figure 10. Erlang model of the first-passage-time distributions. (a) Profile log-likelihood, relative to its maximum, as a function of the effective event number m, for different temperatures at N=1000.*

# Appendix E. Correlation lengths

The spatial correlation lengths are extracted by fitting the low-wavevector behavior of the directly computed structure factors to the Ornstein–Zernike form,

$$S_k(q) = \frac{S_k(0)}{[1 + (q\xi_k)^2]} \qquad (A16)$$

Where $k = conf, iso, 4$. The fits exclude the zero-wavevector mode. The structure factors are computes as

$$S_{conf}(q,t) = \frac{1}{N}\left\langle \left| \sum_i \delta\pi_{ri}(t) e^{i\boldsymbol{q}\cdot\boldsymbol{r}_{ri}} \right|^2 \right\rangle_r \; ; \quad S_{iso}(q,t) = \frac{1}{N}\left\langle \left| \sum_i \delta o_{ri}(t) e^{i\boldsymbol{q}\cdot\boldsymbol{r}_{ri}^{\alpha}} \right|^2 \right\rangle_{r\alpha} \qquad (A17)$$

where $\boldsymbol{r}_{ri}$ denotes the initial position of particle $i$, and an angular average over wavevectors of equal magnitude $q = |\boldsymbol{q}|$ is understood. In addition, $S_4(q,t) = S_{conf}(q,t) + S_{iso}(q,t)$. Not that Eq 16 is a convenient model, as formally, the structure factors cannot all satisfy Eq. A16.

As an independent consistency check, we also fit the corresponding real-space correlation functions to the two-dimensional Ornstein–Zernike form $A\,K_0(r/\xi) + c$, where $K_0(x)$ the modified Bessel function of the second kind of order zero and the offset $c$ is estimated from the long-distance tail. Although the two fitting forms are Fourier-transform pairs in two dimensions, the corresponding estimates are affected differently by finite-size effects, the limited ranges of accessible wavevectors and distances, and deviations from the Ornstein–Zernike form. Agreement between the two approaches and across systems sizes therefore provides evidence that the extracted correlation lengths are robust against these sources of systematic uncertainty.

Figure 11 illustrates the two fitting procedures for $N = 4000$ at $T = 0.45$. The grey shaded regions indicate the fitting intervals: in Fourier space, we restrict the analysis to $0 < q \leq 1$ to probe the long-wavelength behavior, while in real space we fit over $2 \leq r \leq 15$, excluding short-distance microscopic correlations and the long-distance region where statistical noise and uncertainties in the baseline become significant. The close agreement between the resulting correlation lengths supports the robustness of the estimates. Figure 12 reproduces the temperature dependence of all extracted correlation lengths, including that of $\xi_4$.

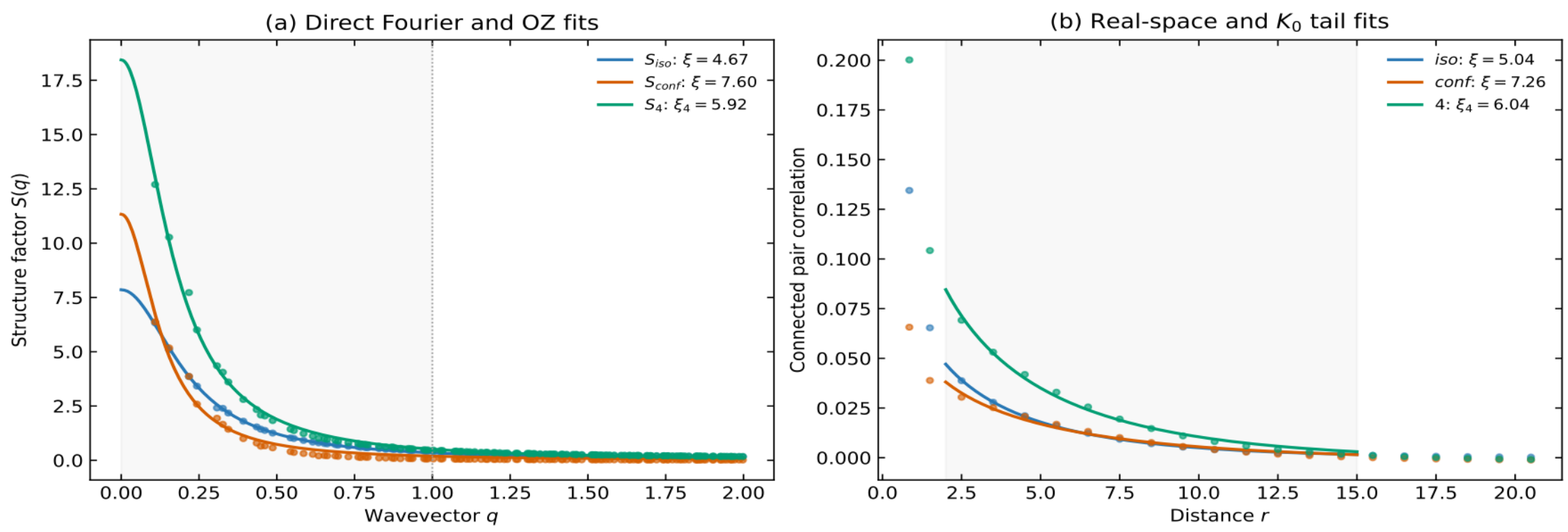

*Figure 11. Extraction of the configurational $\xi_{conf}$ and isoconfigurational $\xi_{iso}$ correlation lengths, and of the length $\xi_4$ associated with the total susceptibility, from (a) direct Fourier-space structure factors and (b) real-space correlation functions, for N=4000 at T=0.45. Symbols represent numerical data, solid lines the corresponding fits, and grey shaded regions the fitting ranges.*

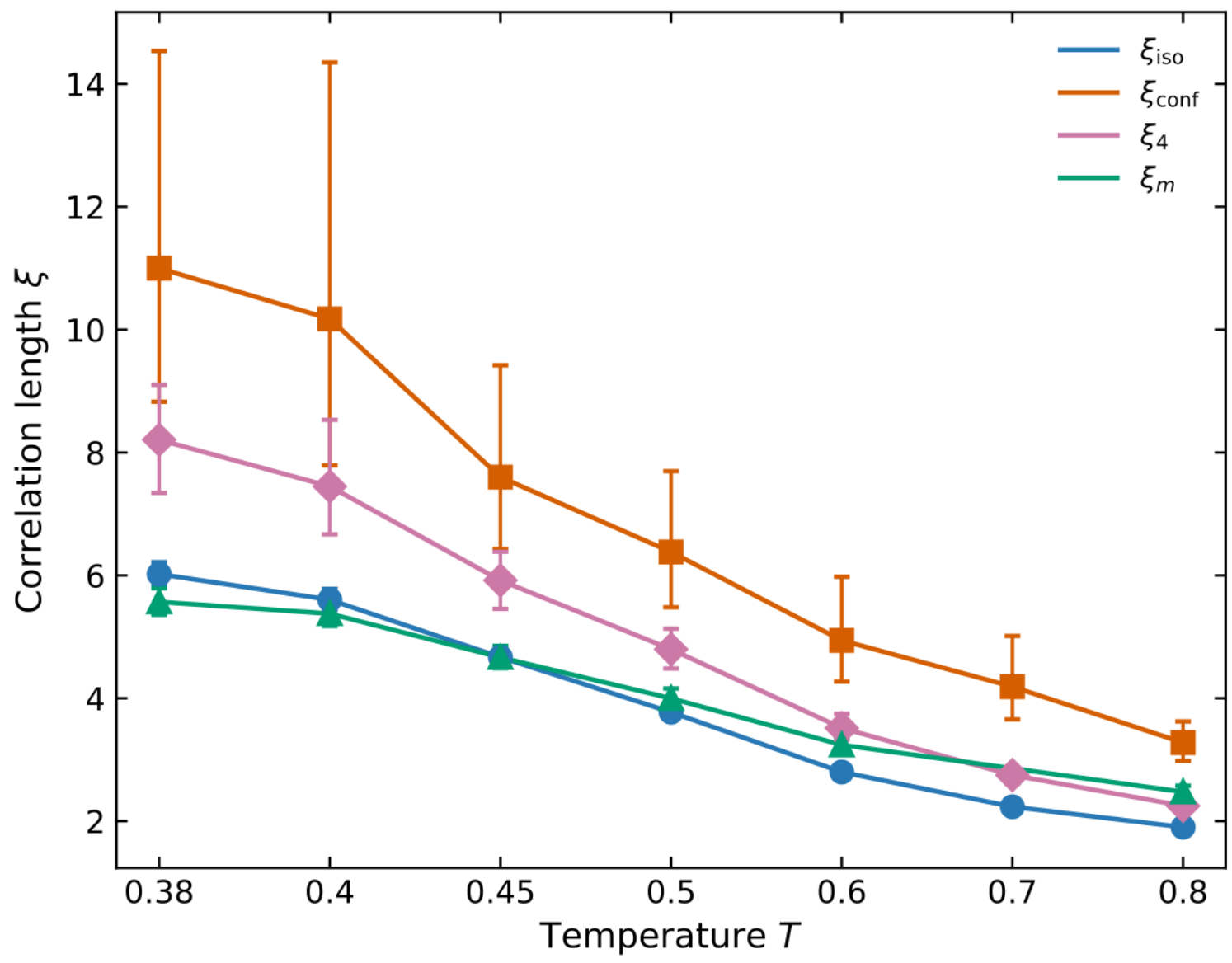

*Figure 12. Temperature dependence of the correlation lengths.*

# Appendix F. Maps of isoconfigurational fluctuations

Figure 13 illustrate maps of the isoconfigurational fluctuating field $\delta o_i^{r\alpha} = o_i^{r\alpha} - \pi_{ri}$ for a fixed configuration $r$ and three different trajectories, $\alpha, \beta$ and $\gamma$. The maps are correlated as the field is small for all particles with a high or a small propensity $\pi_{ri}$. Their difference is concentrated in regions of intermediate propensity.

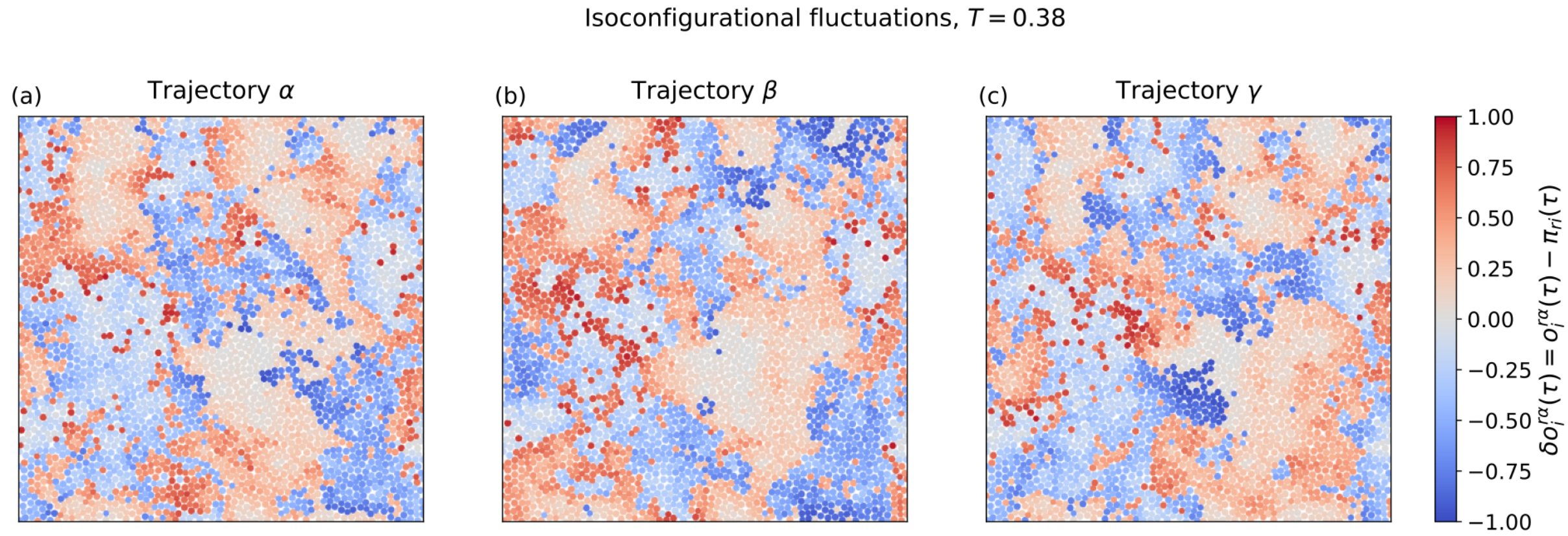

*Figure 13. Maps of the isoconfigurational fluctuating fields $\delta o_i^{r\alpha} = o_i^{r\alpha} - \pi_{ri}$ for a fixed configuration r and three different trajectories, $\alpha, \beta$ and $\gamma$.*